\documentclass[11pt,twoside,A4]{article}
\usepackage{times,fancyhdr}
\usepackage[dvips]{graphicx}
\usepackage{latexsym}
\usepackage[affil-it]{authblk}
\usepackage{mathptm}
\usepackage{amsmath}
\usepackage{amssymb}
\usepackage{color}
\usepackage{setspace}
\usepackage{hyperref}
\usepackage[top=4cm, bottom=4cm, left=4cm, right=4cm]{geometry}

\def\beq{\begin{equation}}
\def\eeq{\end{equation}}
\def\beqn{\begin{eqnarray}}
\def\eeqn{\end{eqnarray}}

\usepackage{slashed}

\newcommand{\be}{\begin{equation}}
\newcommand{\ee}{\end{equation}}
\newcommand{\bea}{\begin{eqnarray}}
\newcommand{\eea}{\end{eqnarray}}
\newcommand{\nn}{\nonumber}

\newcommand{\DIV}[1][\normalsize]{\,\mbox{div}\,}

\begin{document}
\title{Analog Models for Holographic Transport}

\author{Sabine Hossenfelder${}^{1,2}$, Tobias Zingg${}^{1}$}

\affil{${}^{1}$ Nordita, Stockholm University and KTH Royal Institute of Technology\\
Roslagstullsbacken 23, SE-106 91 Stockholm, Sweden\\
${}^{2}$ Frankfurt Institute for Advanced Studies\\Ruth-Moufang-Str.\ 1,  60438 Frankfurt am Main, Germany}

\date{}

\maketitle



\begin{center}
\begin{abstract}
The gauge-gravity duality and analog gravity both relate a condensed matter system to a gravitational theory. This makes it possible to use gravity as an intermediary to establish a relation between two different condensed matter systems: The strongly coupled system from the gauge-gravity duality and the weakly coupled gravitational analog.
We here offer some examples for relations between observables in the two different condensed matter systems. 
In particular, we show that in some cases Green's functions and first order transport coefficients in holographic models are of the same form as those in an analog gravitational system, which allows in principle to obtain the former by measuring the latter.

\end{abstract}
\end{center}

\vskip 11mm

\thispagestyle{empty}

\renewcommand{\thefootnote}{\arabic{footnote}}
\setcounter{footnote}{0}
\setcounter{page}{1}

%
\section{Introduction}
\label{sec:intro}

The AdS/CFT correspondence~\cite{Maldacena:1997re,Witten:1998qj,Gubser:2002tv} relates certain conformal field theories ({\sc CFT}s) with quantum gravitational models that
arise in string theory. The latter, gravitational, theory describes a space-time with negative cosmological constant, the so-called Anti-de Sitter space-time (AdS), while the former
{\sc CFT} is located on the conformal boundary of AdS, i.e.~in a space-time with one spacial dimension less.

The most intriguing feature of the AdS/{\sc CFT} correspondence is that when one of the theories is strongly coupled, then the other is weakly coupled and vice versa. 
Especially useful for practical purposes is the limit when string effects and 
quantum effects are both negligible and the {\sc CFT} is strongly coupled while the gravitational theory is weakly coupled.
In this limit, certain gravitational theories can be used as effective models for strongly coupled condensed matter systems, a technique that
can still be applied when the usual methods of quantum field theory fail. (See e.g.~\cite{Hartnoll:2009sz,McGreevy:2009xe,Sachdev:2010ch} for reviews.)

Intriguingly, there is another approach where a curved space-time emerges in the description of condensed matter systems, and that is analog gravity. Analog gravity is based on the observation that small perturbations around a background medium are described by an equation of motion which is formally identical to that of fields propagating in a curved space-time. Analog gravity hence links weakly coupled gravity with a weakly coupled condensed matter system. 
The relations underlying analog gravity have been known since the mid 1980s~\cite{Unruh:1980cg, Barcelo:2005fc} but only in recent years has the topic begun
to attract attention, quite possibly because experimental realization has become more feasible \cite{Fedichev:2003id,Weinfurtner:2010nu,Steinhauer:2014dra,Steinhauer:2015saa,Euve:2015vml,Peloquin:2015rnl}. 

\begin{figure}
 \centering
 \includegraphics[width=.9\textwidth]{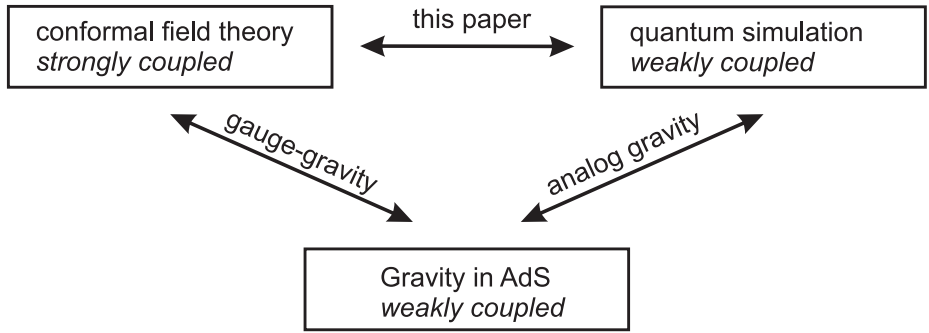}
 \caption{Sketch of relations between the three different systems.}
 \label{fig:sketch}
\end{figure}

Seeing the evident similarities, it lies at hand that one tries to combine both relations -- AdS/{\sc CFT} and analog gravity -- to arrive at
a relation between two condensed matter systems. 
This can be done because AdS/{\sc CFT} and analog gravity both use a curved geometry as an effective description for a condensed matter system,
but the one condensed matter system is strongly coupled and the other one is weakly coupled (see Figure \ref{fig:sketch}). The combination of both relations then -- the ``analog duality'' \cite{Hossenfelder:2014gwa,Hossenfelder:2015pza} -- links a strongly coupled with a weakly coupled condensed matter system.
 
In analog duality, the curved space-time merely constitutes a mathematical framework by help of which computations can be performed. It has no direct correspondence to the actual space-time
geometry in which the condensed matter systems are situated (usually assumed to be flat space).
Nevertheless, this effective geometry can be employed as a mathematical intermediary between the two  different condensed matter systems, which will then, in certain 
aspects, be dual to each other, meaning that quantities in the one description can be identified with quantities in the other description. The purpose of this paper is to present some examples for the, so derived, relations between observables in different condensed matter systems. 

This topic that has been approached from various angles in the last years \cite{Das:2010mk,Semenoff:2012xu,Chen:2012uc,Khveshchenko:2013foa,Bilic:2014dda,Hossenfelder:2014gwa,Hossenfelder:2015pza} and ties into
the overarching scheme of quantum simulation (see e.g.\ \cite{Georgescu:2013oza}). The aim of quantum simulations is to custom-design systems which mimic 
the behavior of a different, mathematically intractable, situation, and then measure the results rather than calculate them. The AdS/{\sc CFT} duality
does not lend itself to this purpose because the gravitational dual does not represent a real-world situation and cannot be simulated in the laboratory. If the gravitational dual, however, is further mapped
onto an analog gravity system, then we are dealing with a correspondence of which both sides are experimentally accessible, thereby making quantum
simulation possible. 

The application that we will focus on in this present work is the calculation of Green functions and first-order transport coefficients. In holography, we have a straightforward procedure to compute these quantities from solutions of a system of partial differential equations in a curved background geometry~\cite{Son:2002sd}.
Mathematically, this is exactly the type of equations which one also has in analog gravity. 

The task thus comes down to finding a pair of an analog gravity system and a holographic dual of another system for which the background geometries are identical. Then the two models have a one-to-one correspondence that maps transport coefficients and Green functions of the holographic model to perturbations of the analog gravity system, provided 
boundary conditions are chosen appropriately. This means any experiment that realizes the suitable analog gravity system can then be employed as an analog dual to compute transport properties in the strongly correlated condensed matter systems with the corresponding holographic dual.

The paper is organized as follows. In order to be self-contained, sections \ref{sec:analogue} and \ref{sec:holo} provide short summaries of the aspects of analog gravity and holography that are relevant for what follows.
In section \ref{sec:examples} we present several examples which illustrate how transport coefficients can be determined by the solution of a scalar field in a curved geometry. In section \ref{sec:generalizations} we discuss how the results presented here can be generalized. 

Units are chosen such that the speed of light and $\hbar=1$. The reader be warned that $c$ denotes the speed of sound in the 
analog model and {\sl not} the speed of light.
Our metric convention is the `mostly plus' signature, $(-1,1,\ldots,1)$.

\section{Analog Gravity}
\label{sec:analogue}

We begin with briefly summarizing the key idea of analog gravity. 
Consider we have a complex scalar field $\phi$ with Lagrangian
\beqn
\mathcal{L} =  \eta^{\mu \nu} \partial_\nu \phi \partial_\mu \phi^* - m^2 \phi \phi^* + V(x,t,\phi,\phi^*)~. \label{Lagphi}
\eeqn
Here and in the following we will allow the potential to have an explicit coordinate dependence because we have in mind interactions that are designed
in the laboratory. Such interactions are typically induced by the presence of external fields \cite{Herring} and are chosen for the very purpose of creating
a specific quantum simulation. We assume that the potential $V$ breaks the global $U(1)$ symmetry in a stable minimum. This
minimum will define our background field.  $\eta$ is the background metric of the space-time in which the field resides and is assumed
 to be the Minkowski-metric.

The complex scalar field $\phi$ can be expressed in terms of two real scalar fields as $\phi = \varphi \exp( i \theta)$, so that the Lagrangian takes the form
\beqn
\mathcal{L} =  \eta^{\mu \nu} \partial_\nu \varphi \partial_\mu \varphi + \varphi^2 \partial_\nu \theta \partial^\nu \theta  - m^2 \varphi^2 + V(x,t,\theta,\varphi)~.
\eeqn
We can then derive the equation of motion for $\varphi$
\beqn
\eta^{\mu \nu} \partial_\nu \partial_\mu \varphi +2 m^2 \varphi = \frac{\partial V}{\partial \varphi} - 2 \varphi \chi ~, \label{eomvarphi}
\eeqn
where 
\bea
\chi &:=& \eta^{\mu\nu} \left( \partial_\nu \theta \right) \left( \partial_\mu \theta \right)
\, 
\eea
is the kinetic term of the phase-field $\theta$. Assuming that the left hand side of (\ref{eomvarphi}) is negligibly small, we
solve this equation to express $\varphi$ as a function of $\chi$ and insert this back into the Lagrangian. This will
generically result in an effective Lagrangian for $\theta$ of the form
\bea
\mathcal{L}_{\theta} &=& \mathcal{L}[\chi(\partial \theta), V(t,x,\theta)]
\, .
\label{eq:Lagranalogian}
\eea
In particular, if the original potential was polynomial in $\phi$, the resulting Lagrangian will contain some (in general fractional)
power of $\chi$.  This is a good effective theory so long as the classical equations of motion are approximately
valid and $\varphi$ is slowly varying. It is a common limit to use in the treatment of Bose--Einstein condensates.

For the analog gravitational system will denote with $\theta_0$ a background field that solves the Euler--Lagrange equations for~\eqref{eq:Lagranalogian} and then consider a fluctuation around this solution,
\bea
\theta	&=&	\theta_0 + \varepsilon \theta_1 + \varepsilon^2 \theta_2 + \ldots
\, .
\eea
Demanding that this expansion again solves the Euler--Lagrange equations leads to equations of motion for the fluctuations $\theta_j$ when expanding in orders of $\varepsilon$ \cite{Barcelo:2005fc}.
Focusing on the lowest order fluctuation, the equation of motion can be brought into the form
\bea
\frac{1}{\sqrt{|g|}} \partial_\mu \left( \sqrt{|g|} g^{\mu \nu} \partial_\nu \theta_1 \right) -   m_{\rm{eff}}^2 \theta_1	&=&	0
\, ,
\label{eq:analog_eom}
\eea
where the (inverse of the) ``acoustic metric'' is defined as
\bea
\sqrt{-g} g^{\mu \nu} &=& -\frac{\partial^2 \mathcal{L}}{\partial (\partial_\nu \theta) \partial (\partial_\mu \theta)} \Bigg|_{\theta=\theta_0} ~,
\label{eq:geff}
\eea
and the ``effective mass'' of the perturbation is
\bea
\sqrt{-g} m_{\rm eff}^2 &=& - \frac{\partial^2 \mathcal{L}}{\partial \theta \partial \theta} + \partial_\nu \left( \frac{\partial^2 \mathcal{L}}{\partial(\partial_\nu \theta) \partial \theta}\right) \Bigg|_{\theta=\theta_0}
\, .
\label{eq:meff}
\eea
The significance of~\eqref{eq:analog_eom} is that this is the exact same equation which also describes the propagation of a field in a curved geometry with metric $g_{\mu\nu}$.
As a consequence, fluctuations in a condensed matter system can be used to simulate physics in curved space-time even though the `real' space-time in which
the system is located remains flat.

Note that if the potential in \eqref{Lagphi} did respect the global U(1) symmetry, then the perturbation $\theta_1$ will be massless, as
is expected for the Goldstone boson. One can, however,
generate masses for $\theta_1$ by explicitly breaking the U(1) symmetry. This corresponds to the familiar `tilt' of
the potential that e.g.~gives masses to pions due to the breaking of chiral symmetry. We will make use of this
explicit symmetry breaking later on. 

For practical purposes, it is often useful to rewrite \eqref{eq:geff} in quantities that are more commonly used for gases and fluids.
For this one notes that the Lagrange density~\eqref{eq:Lagranalogian} defines a stress-energy-tensor,
\bea
T^{\mu \nu} = - \theta^\mu \frac{\partial \mathcal{L}}{\partial (\partial_\nu \theta)} + \mathcal{L} \eta^{\mu \nu}
\, .
\eea
This can be rewritten as the stress stress-energy-tensor of a fluid,
\bea
T_{\mu \nu} = (p_0+\rho_0)u_\mu u_\nu + p_0 \eta_{\mu \nu} ~,
\eea
where four-velocity, pressure, and density of the background field are given by
\bea
u_\nu = \frac{\partial_\nu \theta}{\sqrt{-\chi}} 	\, , \quad
p_0 = \mathcal{L}									\, , \quad
\rho_0 = 2 \chi \frac{\partial \mathcal{L}}{\partial \chi} - \mathcal{L}	\, ,
\label{eq:ident}
\eea
such that the four-velocity is normalized as it should
\bea
\eta^{\mu\nu}u_\mu u_\nu = - 1~.
\eea
The field equations of the relativistic fluid are then identical to the conservation of the stress energy
\bea
\partial_\nu T^{\mu \nu} = 0~,
\eea
and the acoustic metric and its inverse can be expressed as
\bea
g^{\mu \nu} &=& c^{\frac{2}{n-1}}  \left(  \frac{\rho_0+p_0}{-2\chi} \right)^{-\frac{2}{n-1}}  \left( \eta^{\mu\nu} + \left(1- \frac{1}{c^2} \right) u^\mu u^\nu \right) ~, \\
g_{\mu \nu} &=& c^{\frac{-2}{n-1}}  \left(  \frac{\rho_0+p_0}{-2\chi} \right)^{\frac{2}{n-1}}  \left( \eta_{\mu\nu} + \left(1- c^2 \right) u_\mu u_\nu \right) ~.
\label{eq:endderiv}
\eea
Here, $c$ is the speed of sound and defined by $c^{-2} = \partial \rho_0/\partial p_0$.

In the non-relativistic limit one has $p_0 \ll \rho_0$ and $v^2 \ll c^2$ and then the acoustic metric is of the form
\bea
g^{\mu \nu} (t, {\vec x}) &\propto& \left( \frac{\rho_0}{c} \right)^{-\frac{2}{n-1}}
\left( \begin{array}{cc}
-1/c^2 & - v_0^j/c^2 \\
-v_0^i/c^2 & \delta^{ij} - v_0^i v_0^j/c^2  \end{array} \right)~, \label{gupd} \\
g_{\mu \nu} (t, {\vec x}) &\propto& \left( \frac{\rho_0}{c} \right)^{\frac{2}{n-1}}
\left( \begin{array}{cc}
-(c^2-v_0^2) & - (v_0)_j \\
- (v_0)_i & \delta_{ij}   \end{array} \right) \label{gdownd} ~.
\eea
In this limit, the equations of motion for the background field are the familiar continuity equation and the 
Euler equation:
\bea
\partial_t \rho_0 + {\vec \nabla} \cdot (\rho_0 {\vec v}_0 ) &=& 0 \label{continuity} ~, \\
\rho \left[ \partial_t {\vec v}_0 + ( {\vec v}_0 \cdot {\vec \nabla})  {\vec v}_0 \right] &=& \vec{F} \label{euler}~.
\eea
If the fluid is non-viscous, has vanishing rotation (i.e.~is vorticity-free), and is barotropic, we can further simplify this expression. 
The velocity field is then the gradient of a scalar field ${\vec v}_0 = - {\vec \nabla} \phi$ and the density $\rho_0$ is a function of $p_0$ only. 
In this case, the Euler equation can be integrated once and then be written as
\bea
 \partial_t \phi =  h + \frac{1}{2} \left( \vec \nabla \phi \right)^2~, \label{euler3}
\eea
where 
\bea
h(p) = \int_0^p \frac{dp'}{\rho_0(p')} ~
\eea
is the specific enthalpy.

\section{The Holographic Dictionary}
\label{sec:holo}

In this section we will summarize how the AdS/{\sc CFT} correspondence connects observables of the
boundary {\sc CFT} with the gravitational theory in the bulk. In section \ref{sec:examples} we
will then go through these observables again and further connect them with observables of the analog gravity system. 

Since the {\sc CFT} is located on the boundary of the space-time in which the gravitational theory operates, the AdS/{\sc CFT} correspondence is frequently
referred to as ``holographic''. Holography relates a gravitational theory, i.e.~one with dynamical geometry, in $D+1$ dimensions to a non-gravitational field theory in $D$ dimensions.

Because the $D+1$ dimensional bulk is an AdS space-time, it is not globally hyperbolic. This means that initial conditions on a 
space-like slice do not uniquely determine the propagation of fields in this space-time. For a problem to be well-defined, therefore, initial conditions have to 
be supplemented by additional conditions on the conformal boundary when approaching spacial infinity. This means that
the key ingredient of holography is to convert the computation of observables in the strongly coupled
{\sc CFT} on the boundary into a boundary value problem in the curved geometry of the bulk. 

We can picture the field theory dual as `living on the boundary' since its degrees of freedom are given by the asymptotic behavior of the bulk fields.
Schematically, using $\Phi_I$ to denote the collection of all bulk fields, including the geometry, the boundary degrees of freedom are obtained from asymptotic scaling relations of the form
\bea
\phi_I \;\;=\;\; \lim_{z \to \infty} z^{-h_I} \Phi_I	\, ,
\label{eq:holo_fields}
\eea
where the exponent $h_I$ depends on the field and $z$ is the bulk-coordinate in AdS which approaches zero at the boundary.

In the dual field theory, these $\phi_I$ are sources that couple to an operator $\cal{O}^I$,
and their correlation functions are evaluated as
\bea
\left\langle {\cal{O}}^1(x_1) \cdots {\cal{O}}^m(x_m) \right\rangle = (- {\rm i})^{m+1} \left.\frac{\delta^m {\cal{S}}^{{\rm on-shell}}}{\delta \phi_1(x_1) \cdots \delta \phi_m(x_m)} \right|_{\Phi_I \equiv \Phi_I^{(0)}}	\; .
\label{eq:holo_corr}
\eea
Another way to say this is that the gravitational action
\bea
{\cal{S}} \;\;=\;\; \int dx^{D+1}{\cal{L}}_{\rm Bulk}[\partial_\mu \Phi_I, \Phi_I]	\, .
\label{eq:holo_action}
\eea
evaluated ``on-shell'', i.e.~on a solution $\Phi_I^{(0)}$ of the corresponding Euler-Lagrange equations, serves as the generating functional of connected correlators of the field theory dual.
The exponents $h_I$ in~\eqref{eq:holo_fields} are then related to the conformal dimension of the corresponding operators $\cal{O}^I$.

Evaluating~\eqref{eq:holo_corr} may, at first sight, appear uselessly difficult, given that the equations of motion in systems with 
dynamical geometry are highly non-linear and explicit solutions for arbitrary boundary conditions are in general not known analytically.
What makes the formula useful, though, is that it can be evaluated perturbatively. That is, when a specific background $\Phi_I^{(0)}$ is chosen, then 
the $n$-point functions can be calculated by making a perturbative expansion up to order $n-1$.

For this to work we assume that the equations of motion are Hamiltoinian\footnote{Even higher derivative theories can usually be brought to the form~\eqref{eq:holo_action} and having a Hamiltonian functional is generally to be expected in physically realistic systems}, so that a particular solution is entirely characterized by $\{\pi^I, \phi_I\}$, where $\pi^I$ are the conjugate momenta to the fields $\phi_I$.
The momenta can be expressed in terms of boundary values of derivatives of the bulk fields $\Phi_I$, but can as well be identified using the standard relation
\bea
\pi^I \;\;=\;\; \frac{\delta {\cal{S}}^{{\rm on-shell}}}{\delta \phi_I}	\, .
\label{eq:holo_momenta}
\eea
A specific solution to the equation of motion is selected by fixing $\{\pi^I, \phi_I\}$ on a characteristic surface -- usually at the AdS boundary -- which gives rise to relation of the type
\bea
\pi^I \;\;=\;\; \pi^I[\phi_{J}]	\, .
\label{eq:pi_phi_relation}
\eea
Next, one makes a perturbative expansion of the bulk fields around the background solution
\bea
\Phi_I &\longrightarrow & \Phi^{(0)}_I + \varepsilon\Phi^{(1)}_I + \varepsilon^2\Phi^{(2)}_I + \ldots \, .
\eea
This induces an expansion of the momenta in terms of the boundary fields which has the general form
\bea
\pi^I \;\;=\;\; {\pi^I}^{(0)} +  \varepsilon {\cal{G}}^{IJ} \phi^{(1)}_{J} +  \varepsilon^2 {\cal{C}}^{IJK}\phi^{(1)}_{J}\phi^{(1)}_{K} + \ldots	\, .
\label{eq:pi_phi_relation}
\eea
The coefficients in this expansion then correspond to the $n$-point correlation functions from~\eqref{eq:holo_corr}.
The special case of the $2$-point function corresponds to the holographic Green function~$\cal{G}^{IJ}$, which will be discussed further in sections \ref{sec:greens} and \ref{sec:trans}.

The details of the relation~\eqref{eq:pi_phi_relation} depend on the properties of the space of solutions in which one studies a particular problem or configuration. This often means that we must apply further consistency-conditions in
the bulk. Usually, these are condition like demanding that the bulk geometry is smooth or that  curvature singularities are hidden behind a regular event horizon (the latter to ensure that the geometry remains non-singular when time is Wick-rotated to Euclidean signature). In such cases, the consistency conditions in the bulk usually boil down to a set of boundary conditions at the event horizon. 

The most straight-forward aspects of the correspondence relate the metric in the bulk with properties
of the theory on the boundary. Of particular interest are space-times with an event horizon (in the following referred to as black hole space-times) because their duals describe strongly coupled {\sc CFT}s at finite temperature.
 
In such a case, if the AdS-space contains a black hole, then the Hawking temperature associated with the black hole horizon via the surface gravity corresponds to the temperature of the {\sc CFT} on the boundary. The entropy density of both systems is the same. From the metric in the bulk one can further extract the stress-energy-tensor on the boundary with a suitable renormalization procedure that strips off the infinites which the asymptotic limit brings \cite{deHaro:2000vlm}.

\section{The New Dictionary of Analog Duality}
\label{sec:examples}

We here consider a model that has been widely used in the literature as a phenomenological to study electric transport, in particular in holographic superconductors~\cite{Hartnoll:2008vx,Hartnoll:2009sz}.
It consists of a scalar field, $\psi$, which is charged under a $U(1)$ gauge field, $A$, and minimally coupled to Einstein gravity,
\bea
{\cal{S}} 	& = & - \int *\left(R-2\Lambda\right) - \frac{1}{2} \int F \wedge * F \nn \\
		& & \qquad\qquad\qquad- \frac{1}{2} \int \left(d+i \frac{q}{L} A \wedge\right)\psi^{\dagger} \wedge * \left(d-i \frac{q}{L} A \wedge\right)\psi - \frac{m^2}{2 L^2} \int \psi^{\dagger} \wedge *\psi
\, .\qquad\qquad
\label{eq:action_example}
\eea
The equations of motion from this action are a combination of Einstein's field equations, the Maxwell equation and a Klein--Gordon equation for the scalar field,
\bea
G_{\mu\nu} + \Lambda g_{\mu\nu}	&=&	T^{M}_{\mu\nu} + T^{sc}_{\mu\nu}		\, , \label{eq:einstein_eom}\\
\nabla^\mu F_{\mu\nu}			&=&	J_{\nu}		\, ,	 \label{eq:maxwell_eom}\\
\left(\nabla_\mu-i\frac{q}{L}A_\mu\right)^2 \psi - \frac{m^2}{L^2} \psi			&=&	0  \label{eq:scalar_eom}	\, .
\eea
With the usual stress-energy tensors for the field stength and a charged scalar, respectively,
\bea
T^{M}_{\mu\nu}	&=&	F_{\mu\kappa} {F_{\nu}}^{\kappa} -\frac{F^2}{4} g_{\mu\nu}	\, ,	\\
T^{sc}_{\mu\nu}	&=&	\frac{1}{2}\left(\nabla_\mu+\frac{iq}{L}A_\mu\right)\psi^\dagger \left(\nabla_\nu-\frac{iq}{L}A_\nu\right)\psi -\frac{1}{4} g_{\mu\nu}\left[ |(\nabla_\mu-iqA_\mu) \psi|^2 +  \frac{m^2}{L^2} |\psi|^2  \right]	\, ,\qquad
\eea
and the current
\bea
J_{\nu}	&=&	\frac{i q}{L} \left( \psi^\dagger \nabla_\nu \psi - \nabla_\nu \psi^\dagger \psi \right) \, .
\eea
A frequently studied background solution of these equations are time-independent space-time geometries with translational symmetry in the transverse directions, such that the metric can be parametrized as
\bea
\frac{ds^2}{L^2} &=& \frac{{\rm e}^{2 C(z)}}{z^2} \left[ - f(z) dt^2 - 2\sqrt{1-f(z)} dt\,dz + dz^2 + dx^2 + dy^2 \right]
\, ,
\label{eq:bgmetric}
\eea
where $L$ is the typical length scale in the bulk and the coordinate $z$ goes to zero on the AdS conformal boundary. For later convenience, we also introduce a frame, $e^i$, adapted to the diagonal form of the metric,
\bea
\frac{e^0}{L}	\;=\;	\frac{{\rm e}^{C} \sqrt{f} }{z} \left[ \, dt + \frac{\sqrt{1-f}}{f}\, dz\right]	\; , \quad
\frac{e^1}{L}	\;=\;	\frac{{\rm e}^{C}}{z} dx	\; , \quad
\frac{e^2}{L}	\;=\;	\frac{{\rm e}^{C}}{z} dy	\; , \quad
\frac{e^3}{L}	\;=\;	\frac{{\rm e}^{C}}{z \sqrt{f}} dt	\; . \quad
\label{eq:frame}
\eea
In this case $\psi$ is also a function of $z$ olny, $\psi = \psi(z)$, and the gauge field is of the form
\bea
A = L\,a(z) \left[ \, dt + \frac{\sqrt{1-f}}{f}\, dz\right]	\, , \qquad
F = L\,b(z) \, dz \wedge dt	\, .
\label{eq:bggauge}
\eea
Explicit expressions for $C, f, a, \psi$ in closed form are only known for $\psi \equiv 0$, when the metric reduces to the AdS Reissner-Nordstr{\"o}m space-time, which is given by
\bea
f(z) 	\;=\; 1 - (Q^2+1)z^3 + Q^2 z^4	\, , \quad
C(z) 	\;=\; 1	\, ,	\quad
a(z)	\;=\; Q(1-z) \, .
\label{eq:RNsolution}
\eea
In the following we focus mostly on the simple action~\eqref{eq:action_example} as an illustrative example, we wish to emphasize that it was shown in~\cite{Hossenfelder:2017iom} that any metric of the form~\eqref{eq:bgmetric} can be realized as an analog metric.

\subsection{Background values}
\label{sec:bg-values}

To illustrate the general idea, we will start with some simple examples that build on the previous works \cite{Hossenfelder:2014gwa,Hossenfelder:2015pza}. 
As these papers show, one of the metrics most commonly used for holographic models, that of a (charged) planar black hole in asymptotic AdS, has
an analog dual even without introducing a conformal prefactor. We will denote the position of the horizon in direction $z$ as $z_0$. 

In the non-relativistic limit and $D=4$, the speed of sound, $c$, of the background fluid is constant  and the energy-density and velocity-field are given by
\beqn
\rho_0 = c m^2 a^2 \frac{L^2}{z^2}~,~ v_0 = \sqrt{c} \frac{z}{z_0}~, \label{bgvals}
\eeqn
where $m$ is the mass of the particles which the analog fluid is composed of and $a$ is a parameter of dimension mass that quantifies the overall amplitude of fluctuations. The velocity field points into the direction perpendicular to the
black hole horizon.

On the other hand, we know that the temperature of the condensed matter system on the AdS boundary is given by
\beqn
T = \frac{1}{\pi z_0 L}~,
\eeqn
from which we obtain the relation
\beqn
\frac{1}{\sqrt{c}}\partial_z v_0 \Big|_{z=z_0} = \frac{T}{L}~, \label{partzvo}
\eeqn
where the quantities on the left side belong to the weakly coupled analog gravity system, while those on the right side belong to the strongly coupled {\sc CFT}. 

This relation in and by itself does not provide much insight. It merely tells us which parameter in the one
system belongs to which parameter in the other system, and is therefore necessary for a quantitative
comparison. However, to better understand the systems themselves, we want to have relations between
quantities on the one side corresponding to relations on the other side. 

We can further extract the stress-energy-tensor of the {\sc CFT} from the metric, from which one obtains 
\beqn
\langle T_{\nu}^\kappa \rangle = \pi z_0^2/L^3 {\rm{diag}}(-3,1,1,1)~.
\eeqn
By using  (\ref{partzvo}) one can then express the energy-density of the {\sc CFT} through the gradient of the velocity field of the fluid. 

\subsection{Green functions}
\label{sec:greens}

Of special interest for any field theory are the Green functions or propagators, respectively.
They are of central importance because the Green functions are directly related to measurable quantities like decay rates, cross sections, and transport coefficients.

To compute the Green functions, we use the common method of linear response \cite{Son:2002sd}. By virtue of the holographic dictionary, the Green functions of the boundary~{\sc CFT} can be evaluated by solving the equations of motion for perturbations of the corresponding bulk fields with appropriately chosen boundary conditions. The Green functions in the bulk themselves are calculated by considering infinitesimal perturbations around the metric (\ref{eq:bgmetric}) and the fields on it (\ref{eq:bggauge}) so that the equations of motion are still satisfied. 
These resulting equations describe the propagation of these perturbations through the space-time~\eqref{eq:bgmetric}. 
This method therefore results in a Klein--Gordon equation structurally identical to the
equations of perturbations in analog gravity \eqref{eq:analog_eom}.

Thus, properties of the fundamental degrees of freedom on either side -- correlation functions of an operator in the holographic field theory and sound propagation in the analog fluid -- directly correspond to each other through a relation that is mediated by the bulk space-time.
The geometry translates different aspects of the two condensed matter systems into each other; the systems are analog duals of each other.

However, the equations for a generic perturbation in a gravitational system will lead to a large system of coupled, partial, differential equations. This makes the search for an analog model that leads to the same equations a quite daunting task.

Luckily, there are some interesting cases in which these equations decouple, thereby greatly simplifying the calculation. In these cases, the Green functions can then be computed using a perturbation $\delta \Phi$ of only a single scalar field which satisfies an equation of motion of the general form
\bea
\square \delta \Phi - \widetilde m^2_{\rm eff} \delta \Phi &=& 0
\, .
\label{eq:scalar_eq}
\eea
Here, $\square$ is the d'Alembertian of the effective bulk background metric $g$, and 
$\widetilde m_{\rm eff}$ is an effective mass-term. 

To establish an analog duality we will therefore have to generate the particular mass $\widetilde m_{\rm eff}$ without altering
the background metric. We could write down the equations that derive from this, but solving them will not in general be possible. We will
instead look at a particular example for the potential to illustrate how it works. For this, we will use the case
previously discussed in \cite{Hossenfelder:2014gwa}. As shown in this previous work, for the case of the planar black hole
in 4+1 AdS space, the continuity equation works out to be just
\beqn
\partial_z (\rho_0 v_0) = 0 ~.
\eeqn
The Euler-equation then allows one to calculate the force-density $F_z$ necessary to get the
required pressure. In the static case it takes the form
\beqn
\partial_z (\rho_0 (v_0)_z^2) + c \partial_z p_0 = - F_z = \partial_z V ~.
\eeqn
Using 
\beqn
\partial_z p_0 = \partial_z \rho_0 \frac{\partial p_0}{\partial \rho_0} = c \partial_z \rho_0~,
\eeqn
one obtains
\beqn
F_z = 2 c^2 (Lma)^2 \frac{\gamma(z)}{z^3}~. \label{Fz}
\eeqn
Let us now suppose we have a potential of the form
\beqn
V(z,\theta) = a_1(z) \theta^2  - a_2 (z) \theta^4~,
\eeqn
which generates the background solution $\theta_0$. 
Since we know the velocity-profile we can integrate it to get the field, so we know the coordinate-dependence
of the entire potential. In the case under consideration that is
\beqn
\theta_0 = {\rm const.} + \int dz (v_0)_z = {\rm const.} + \frac{\sqrt{\kappa}}{z_0} z^2~.
\eeqn
One can then use \eqref{Fz} to find suitable functions $a_1(z)$ and $a_2(z)$. But from \eqref{eq:meff} we
further have
\beqn
m_{\rm eff}^2 = {z^2} \left( 2 a_1 (z) -12 a_2(z) \theta_0^2 \right)~,
\eeqn
which will in general not be the correct effective mass to probe the Green function.
Our task is then to find a new potential,
\beqn
\widetilde{V}(z,\theta) = \widetilde{a}_1(z) \theta^2  - \widetilde{a}_2 (z) \theta^4~, \label{tildeV}
\eeqn
which still generates the background solution $\theta_0$ but changes the effective mass $m_{\rm eff}$ to the
$\widetilde m_{\rm eff}$  necessary to obtain the equation of motion \eqref{eq:scalar_eq} for the 
perturbation. 
This leads to the requirements
\beqn
\frac{\partial V}{\partial \theta} \Big|_{\theta=\theta_0} = \frac{\partial {\widetilde V}}{\partial \theta} \Big|_{\theta=\theta_0} ~,~
\frac{1}{\sqrt{-g}}\frac{\partial^2 {\widetilde V}}{\partial \theta^2} \Big|_{\theta=\theta_0} = \widetilde m^2_{\rm eff} ~.
\eeqn
These equations can be solved to give
\beqn
\widetilde a_1(z) = \frac{z^2 (m_{\rm eff}^2 - \widetilde m_{\rm eff}^2)}{8} + a_1(z) ~,~\widetilde a_2(z) = \frac{z^2(\widetilde m_{\rm eff}^2 -  m_{\rm eff}^2)}{8 \theta^2_0} + a_2(z)~.
\eeqn
By this we have expressed $\widetilde a_1(z)$ and $\widetilde a_2(z)$ entirely through functions whose $z$-dependence is known already due to the requirements on the background fluid and the necessity to reproduce the effective mass.

This change of the potential will, of course, change the equations of motion in general, but in such a way that it is solved by the particular solution~$\theta_0$. The analog acoustic metric~\eqref{eq:geff} for a perturbation will also remain unchanged. What will change is only the effective mass of this perturbation. One sees from the simple example given above that this is generally possible, provided the
potential has at least two interaction terms, so that the minimum can be kept while the second derivative at the minimum changes.

Ultimately, it will be the experimental possibilities that determine which $z$-dependence of the interaction can be realized, and thus, which types of perturbations and Green functions can be simulated. Nevertheless, the calculation presented here shows that
rather simple adjustment of couplings allow to make a connection to the Green functions in the dual system.

\subsection{Linear Response and Transport Coefficients}
\label{sec:trans}

Transport coefficients play an important role in relating theoretical results to experiment.
They measure how rapidly a perturbed system is returning to equilibrium and can thus be directly related to data gathered from measurements.
These coefficients are intimately related with Green functions by an equation known as the Kubo formula.

These quantities are related as follows. Suppose we have a system in an equilibrium state $\Phi_J$, and make a small perturbation, $\delta\Phi_J$, away from equilibrium. The Green function ${\cal{G}}^{IJ}$ is defined
as the function which encodes the response $\delta\Pi^I = {\cal{G}}^{IJ} \cdot \delta\Phi_J$. 
The corresponding transport coefficients, call them $\gamma^{IJ}$, can then be expressed in the form
\bea
\gamma^{IJ} \;\; \propto \;\; \lim\limits_{\omega \to 0} \frac{{\cal{G}}^{IJ}(\omega,0)}{i \omega}
\, ,
\label{eq:Kubo}
\eea
where $\omega$ is the frequency of the perturbation.
In this way, we can extract transport coefficients in holographic models by studying linear response around a given background space-time \cite{Herzog:2011ec}.

Transport coefficients that are often considered in holographic models are the electrical conductivity (calculated from the response to changes in the applied electric field) and shear viscosity (calculated from the response to applied transversal shear). In the following subsections, we will use these as examples to demonstrate how a dictionary can be established between the strongly coupled {\sc CFT} and a condensed matter system with an analog gravitational description.

\subsubsection{The scalar $2$-point function}

We will begin with the simplest case, the $2$-point function $\langle \cal{O} \cal{O} \rangle$ of a scalar operator $\cal{O}$ on the boundary that is dual to a scalar bulk field.
Such an operator could for example describe a mass density, a charge density, or an order parameter of a phase transition. A scalar $2$-point function is also used to study correlations in the Hawking-radiation in curved space-times which have recently attracted attention \cite{Steinhauer:2015saa,deNova:2018rld}.

In general, perturbing such a scalar field will induce a response in the metric $g$ and, if the field is charged, in the corresponding gauge field.
However, for the case of the scalar field $\psi$ in \eqref{eq:action_example} this complication is avoided when one studies the probe limit or a background with $\psi_0=0$. In particular, for the Reissner--Nordstr\"om solution~\eqref{eq:RNsolution} the back-reaction is quadratic in the perturbation $\delta \psi$ and therefore does not contribute to the $2$-point function obtained from linear response. The equation of motion for the scalar-field perturbation $\delta \psi$ in this background is then simply the Klein--Gordon equation~\eqref{eq:scalar_eom} in this background. From this one can read off the effective mass for the pertubation
\bea
\widetilde m_{\rm eff}^2	&=&	\frac{1}{L^2} \left( m^2 - \frac{q^2 Q^2 z^2 (1-z)^2 }{1 -(1+Q^2)z^3 + Q^2 z^4} \right)
\, , \label{kgeff}
\eea
where $m$ is the mass of the scalar field, $q$ is the U(1)-charge of the scalar field, and $Q$ is the electric charge density of the background. Using the procedure laid out in section \ref{sec:greens}, one can then adjust the potential to generate
the desired effective mass of the perturbation.

\subsubsection{Conductivity}
\label{sec:conductivity}

The electrical conductivity tensor, $\sigma$, is a measure of a material's ability to conduct an electric current. It can be calculated by the Kubo formula
\bea
\sigma^{ij} \;\;=\;\; \lim\limits_{\omega \to 0} \frac{{\cal{G}}_{{\rm em}}^{ij}(\omega,0)}{{\rm i} \omega}
\, .
\eea
Here, the electromagnetic Green function ${\cal{G}}_{{\rm em}}^{ij}$ is defined through the correlator of the electric current density, ${\cal{J}}$, in the holographic dual on the boundary: 
\bea
\langle {\cal{J}}^i \ {\cal{J}}^j \rangle
\, .
\eea
To calculate the Green function via linear response, we study the response of the current $\delta \cal{J}^\nu$ to a fluctuation of the boundary value of the bulk gauge field $\delta A_\mu$. The components of the
Green function are then defined through the relation
\bea
\delta {\cal{J}}^\nu \;\;=\;\; {\cal{G}}^{\nu\kappa}_{{\rm em}} \,\delta A_\kappa
\, ,
\label{eq:Greens_function_em}
\eea
This perturbation of the current on the boundary is, essentially, the normal component of the bulk field strength and thus, in appropriately chosen coordinates, identified via
\bea
\delta F_{\mu\nu} &=& e^3_{[\mu} \delta \cal{J}_{\nu]}
\, ,
\eea
In general, $\sigma^{ij}$ is a two-tensor.
However, in the absence of a magnetic flux, the off-diagonal components -- like the Hall conductivity -- vanish. 
Furthermore, when restricted to the case of vanishing spatial momentum in the transversal direction, there is only one independent component left in the conductivity tensor, which can be chosen as the usual longitudinal conductivity.
This can described by the response to a temporally modulated perturbation. We will use here
a perturbation in the $x$-component of the gauge field $A$.

Due to coupling to the metric, this will also require to add a perturbation in an off-diagonal metric component because otherwise we would not obtain a closed and consistent system of equations.
Since we here consider vector and tensor fields it is convenient to use the background frame~\eqref{eq:frame} to parametrize the perturbations of relevance in this situation
\bea
\delta A_{\mu}	\;\;=\;\; \alpha(t, z) \, e^1_{\mu}
\, , \qquad
\delta g_{\mu\nu} 	\;\;=\;\;	\sqrt{f} \beta(t, z) \, e^0_{(\mu} e^1_{\nu)}
\, .
\label{eq:gauge_perturbation}
\eea
With the choice of parametrization~\eqref{eq:gauge_perturbation}, the functions $\alpha$ and $\beta$ can be related to a single function~$\theta(t, z)$
\bea
\theta	\;\;=\;\; \dot{\alpha}	\, , \qquad
2 z b {\rm e}^{-C} \theta \;\;=\;\; \dot{\beta}' + \frac{\sqrt{1-f}}{f} \ddot{\beta}
\, ,
\eea
where prime and dot denote derivatives with respect to $z$ and $t$, respectively.
The scalar function~$\theta$ is itself a solution to the Klein--Gordon equation with effective mass
\bea
\widetilde m_{\rm eff}^2	&=&
\frac{1}{L^2} \left[
	3 - \frac{(m^2+q^2)\psi^2}{2}
	- {\rm e}^{2C} \left( 1 - z C' \right)
	- 5 {\rm e}^{-4 C} z^4 b^2
	\right]
\, .
\eea
The functions $C$, $b$ and $\psi$ have to be determined from solving the background equations of motion, i.e.~the combined Einstein-Maxwell-scalar equations~(\ref{eq:einstein_eom}-\ref{eq:scalar_eom}).
Explicit expressions are in general not known in analytic form, but it is straightforward to obtain their profile via numerical integration.
One can then again express the requirements on the effective mass as a requirement on the potential, as discussed in
section \ref{sec:greens}.

\subsubsection{Shear Viscosity}

The shear viscosity, $\eta$, can be extracted from a response to a perturbation of the metric
\bea
\eta \;\;=\;\; \lim\limits_{\omega \to 0} \frac{{\cal{G}}^{xy,xy}(\omega,0)}{{\rm i} \omega}
\, ,
\eea
where $\cal{G}^{\nu\mu,\kappa\lambda}$ is the Green function for the response in the holographic stress-energy tensor~${\cal{T}}^{\mu\nu}$ when the boundary metric~$g_{\kappa\lambda}$ is perturbed,
ie it corresponds to the $2$-point function
\bea
\langle \cal{T}^{\mu\nu} {\cal{T}}^{\kappa\lambda} \rangle
\, .
\label{eq:Greens_function_se}
\eea
Using linear response and the standard holographic calculation procedure \cite{Son:2002sd}, it can be extracted from the relation
\bea
\delta {\cal{T}}^{\mu\nu} \;\;=\;\; {\cal{G}}^{\nu\mu,\kappa\lambda} \cdot \delta g_{\kappa\lambda}
\, .
\label{eq:Greens_function_se}
\eea
In the specific case of using the background~\eqref{eq:bgmetric}, the metric is translationally and rotationally invariant in the $xy$-direction and does not couple to any tensor fields. One then expects perturbations in the $g_{xy}$-component to decouple.
Indeed, when we consider a metric perturbation of the form
\bea
\delta g_{\mu\nu}	&=& g(t, z) \, e^1_{(\mu} e^2_{\nu)}
\, ,
\eea
which is parametrized via a single scalar function $g$ and the background frame~\eqref{eq:frame}, then the resulting equation for $\theta$ is a Klein--Gordon equation with effective mass equal to zero. Once again one can then amend the potential
as discussed in \ref{sec:greens} to ensure that the perturbations are of this type. 

We also want to mention that a minimal coupling that breaks isotropy in the $xy$-direction,~e.g.~due to spatially modulated backgrounds~\cite{Withers:2014sja} would create an effective mass for this perturbation. To simplify calculations, such
backgrounds can be approximated with $Q$-lattices~\cite{Donos:2013eha}. A different way to study massless perturbations was recently proposed in \cite{Datta:2018rxy}.

\section{Generalizations}
\label{sec:generalizations}

In the previous section we provided several examples where a transport property of a system with a holographic dual can be equivalently described by an analog model depending on just one scalar field.
For background like~\eqref{eq:bgmetric} with a high degree of symmetry this duality is simple to achieve, and also for the somewhat more complicated cases discussed in~sec.~\ref{sec:examples}, perturbations can be separated into a set of decoupled master equations and
the duality can still be achieved.
In general, however, we expect that a generic perturbation is only consistently described by a system of several coupled partial differential equations, which makes the situation far more complicated. 
Constructing a particular example of an analog model for holographic transport in the general case will be left for future work. But we here want to argue that, besides the logistical difficulty that stems from dealing with an increased number of degrees of freedom, there are no other conceptual obstructions to find an analog model.

To see this, let us first extend the discussion from sec.~\ref{sec:analogue} to a case where the Langrangian depends on more than one, say $N$, fields, i.e.~consider $\cal{L}[\theta^I_{;\mu},\theta^I]$.
For notational purposes, let $\Theta:=\left\{\theta^I\right\}_{I = 1}^{N}$.
Again, perturbatively expanding around a background,
\bea
\Theta	&=&	\Theta_0 + \varepsilon \Theta_1 + \varepsilon^2 \Theta_2 + \ldots
\, .
\eea
This can now be plugged into the corresponding action,
\bea
S[\Theta]	& = &	S[\Theta_0]
			+ \varepsilon\int EL[\Theta_0]\cdot \Theta_1 \sqrt{-g}\,d^nx 
			+ \varepsilon^2\int EL[\Theta_0]\cdot \Theta_2 \sqrt{-g}\,d^nx	\nn \\
	&	&	\qquad\qquad\qquad\qquad
			+ \frac{\varepsilon^2}{2}\int \left( \nabla_\mu\Theta_1^\dagger \cdot G^{\mu\nu} \cdot \nabla_\nu\Theta_1
				+ \Theta_1^\dagger \cdot M \cdot \Theta_1 \right) \sqrt{-g}\,d^nx 
\, . \quad
\eea
Hereby, $EL[\Theta_0]$ denote the Euler -- Lagrange equations for the Lagrangian $\cal{L}$ and, furthermore, the $N \times N$ matrices\footnote{for fixed $\mu, \nu$} $G^{\mu\nu}[\Theta_0]$ and $M[\Theta_0]$ were introduced with entries
\bea
G^{\mu\nu}_{IJ}[\Theta_0]	\;\;=\;\;	\left.\frac{\partial^2 \cal{L}}{\partial \theta^{(I}_{;\mu} \partial \theta^{J)}_{;\nu}}	\right|_{\Theta=\Theta_0}	\, ,	\qquad
M_{IJ}[\Theta_0]	\;\;=\;\;	\left.\frac{\partial^2 \cal{L}}{\partial \theta^{I} \partial \theta^{J}} - 2\nabla_\mu \frac{\partial^2 \cal{L}}{\partial \theta^{(I}_{;\mu} \partial \theta^{J)}_{\color{white}{;\mu}} }	\right|_{\Theta=\Theta_0}	\, .
\eea
			
With the assumption that the background $\Theta_0$ is a solution of the Euler -- Lagrange equations it follows that the equations of motion up to order $\varepsilon^2$ are satisfied if the perturbations $\theta^J_1$ form a solution of
\bea
\nabla_\mu \left(G_{IJ}^{\mu\nu}[\Theta_0]\, \nabla_\nu\theta^J_1\right)
				- M_{IJ}[\Theta_0]\, \theta^J_1	&=&	0
\, .
\label{eq:pert_general}
\eea	
It will now be shown that a set of equations in exactly this same form will result when the metric is made dynamical and the action is minimally coupled to an Einstein -- Hilbert term, if perturbations are parametrized in a convenient way.

As already discussed in sec.~\ref{sec:examples}, when a $(p,q)$ tensor is minimally coupled to gravity, the equations of motion for a perturbation in this degree of freedom can be cast as a set of coupled scalar partial differential equations by choosing a covariantly constant frame $e_{a}^{\mu}$ and parametrizing the perturbation as
\bea
\delta { {\cal{T}}^{\mu_1 \cdots \mu_p}}_{\nu_1\cdots \nu_q}
	&=&	\varepsilon\, {t^{a_1\cdots a_p}}_{b_1\cdots b_q} \,e_{a_1}^{\mu_1} \cdots e_{a_p}^{\mu_1} e^{b_1}_{\nu_1} \cdots e^{a_q}_{\nu_1}
\, .
\eea
Thus, the main caveat to directly conclude that a form like~\eqref{eq:pert_general} must result is because of the curvature term $\int *R$.
In that form, it depends not only on first, but also on second order derivatives of the metric, respectively the frame.
It is however known, that these can be reorganized by adding a total derivative to the action, and an equivalent, though somewhat more convoluted, way to write the Einstein -- Hilbert Lagrangian would be
\bea
\widetilde{\cal{L}}^{EH}	&=&
- de^a \wedge e^b \wedge * (de_b \wedge e_a) + \frac{1}{2} de^a \wedge e_a \wedge * (de^b \wedge e_b) 
\, .
\label{eq:EH_alt}
\eea
With this and parametrizing the perturbation of the frame 
\bea
\delta e_a^\mu
	&=&	\varepsilon{\Omega_a}^b e_{b}^{\mu}
\, ,
\eea
the action can, effectively, be written as a functional depending on the fields $\{\Omega,t\}$ parametrizing the displacement from a given, fixed, background, with the former entering with at most first derivatives.
It is then straightforward, though likely rather tedious, to find equations of motion for the first order perturbation that is exactly of the form~\eqref{eq:pert_general}.
In practice, of course, some of these can be expected to be trivially satisfied, since this form of parametrization will have some redundancies, given that the action is to remain invariant under diffeomorphisms or Lorentz boosts of the frame.

This simplifies the necessary amount of computation compared to higher order correlation functions significantly, as the latter would, in addition, also require to evaluate various Witten diagrams~\cite{Witten:1998qj}.

\section{Conclusions}
\label{sec:blabla}

We have demonstrated here how a correspondence can be established between two seemingly unrelated condensed matter systems by combining the AdS/{\sc CFT} duality with analog gravity.
The key reason why this correspondence holds is that two phenomena -- transport in the strongly coupled system and the propagation of perturbations in the weakly coupled system -- are described by equations with identical mathematical structure.

The relations derived here offer the possibility that future experiments on weakly coupled condensates
may be used to explore the behavior of strongly coupled systems. In particular, as strange metals are presently
believed to be `strange' because they do not have quasi-particles, the link explored here may
turn out useful to better understand such metals by studying the behavior of their analog duals.

An entirely different, maybe even more exciting, application of this work would be to experimentally
test the AdS/{\sc CFT} correspondence (or at least its suitability for the systems under consideration). This becomes possible because
we assumed this, still  unproved, correspondence to be correct. Hence, if it was possible to measure
the properties of two condensed matter systems that are linked in the way discussed here, this would
implicitly prove the validity of the holographic dictionary. 

\section{Acknowledgements}

TZ acknowledges support from Vetenskapsradet under the project number 2015-04852. SH acknowledges support from the German Research Foundation.

\bibliography{Analogbib}
\bibliographystyle{h-physrev5}

\end{document}